\documentclass{PoS}
\newcommand{\aap}{A\&A}
\newcommand{\ap}{Astr. Phys.}
\newcommand{\apjs}{ApJS}
\newcommand{\apj}{ApJ}
\newcommand{\mnras}{Mon. Not. R. Astron. Soc.}
\usepackage{multirow}

\title{Real-Time Analysis sensitivity evaluation of the Cherenkov Telescope Array}

\ShortTitle{CTA/RTA sensitivity evaluation}

\author{\speaker{Valentina Fioretti}$^{,a}$, Andrea Bulgarelli$^{a}$, Andrea Zoli$^{a}$, Sera Markoff$^{b}$, Marc Ribo$^{c}$, Susumu Inoue$^{d}$, Paola Grandi$^{a}$, Giovanni De Cesare$^{a}$, for the CTA Consortium\footnote{Full consortium author list at http://cta-observatory.org}\\
     \llap{$^a$}INAF/IASF Bologna, Bologna, Italy\\
     \llap{$^b$}University of Amsterdam, Amsterdam, The Netherlands\\
     \llap{$^c$}Institut de Ci\`ences del Cosmos, Universitat de Barcelona, Barcelona, Spain\\
     \llap{$^d$}Institute for Cosmic Ray Research, University of Tokyo, Tokio, Japan\\
     E-mail:  \email{fioretti@iasfbo.inaf.it}}

\abstract{The Cherenkov Telescope Array (CTA), the new generation very high-energy gamma-ray observatory, will improve the flux sensitivity of the current Cherenkov telescopes by an order of magnitude over a continuous range from about 10 GeV to above 100 TeV. With tens of telescopes distributed in the Northern and Southern hemispheres, the large effective area and field of view coupled with the fast pointing capability make CTA a crucial instrument for the detection and understanding of the physics of transient, short-timescale variability phenomena (e.g. Gamma-Ray Bursts, Active Galactic Nuclei, gamma-ray binaries, serendipitous sources). The key CTA system for the fast identification of flaring events is the Real-Time Analysis (RTA) pipeline, a science alert system that will automatically detect and generate science alerts with a maximum latency of 30 seconds with respect to the triggering event collection and ensure fast communication to/from the astrophysics community. According to the CTA design requirements, the RTA search for a true transient event should be performed on multiple time scales (from minutes to hours) with a sensitivity not worse than three times the nominal CTA sensitivity.
\\
Given the CTA requirement constraints on the RTA efficiency and the fast response ability demanded by the transient science, we perform a preliminary evaluation of the RTA sensitivity as a function of the CTA high-level technical performance (e.g. effective area, point spread function) and the observing time. This preliminary approach allows the exploration of the complex parameter space defined by the scientific and technological requirements, with the aim of defining the feasibility range of the input parameters and the minimum background rejection capability of the RTA pipeline. 
}

\FullConference{The 34th International Cosmic Ray Conference,\\
		30 July- 6 August, 2015\\
		The Hague, The Netherlands}

\begin{document}

\section{The Real-Time Analysis in the CTA context}

The observation of the Very High Energy (VHE) sky with the current Imaging Air Cherenkov Telescopes (IACTs, e.g. H.E.S.S. \cite{2006A&A...457..899A}, MAGIC \cite{2006A&A...457..899A}, and VERITAS \cite{2002APh....17..221W}) opened a new discovery window on the most energetic phenomena of the Universe and revealed a rich and complex zoo of sources still to be fully understood. At the time of writing, about 160 sources have been discovered to emit in the TeV energy range (see the TeVCat catalog\footnote{\href{http://tevcat.uchicago.edu}{http://tevcat.uchicago.edu}}), comprising among others supernova remnants, pulsars and pulsars wind nebulae, binary systems, and blazars. 
Despite the enormous breakthrough in the study of the non-thermal sky, the present view is just the tip of the iceberg. As a comparison, about 3000 sources are listed in the 4-year catalog \cite{3FGL} of the space gamma-ray Fermi/LAT instrument. Deeper and faster observations are needed to both better understand the physics beyond the VHE sources and catch new unexpected objects (about 30\% of Fermi and AGILE sources have no identified counterpart and are still unkown).
\\
The Cherenkov Telescope Array (CTA, \cite{CTA1}), the new generation very high-energy gamma-ray observatory, will dramatically extend the capability of the current ground-based VHE experiments by building tens of atmospheric Cherenkov telescopes in the Northern and Southern hemispheres. Three telescopes sizes are foreseen, with the few large telescopes to observe the low energy range and many small ones to extend the observational window up to hundreds of TeV.
Thanks to the wide and continuous energy coverage (about four orders of magnitude, from 30 GeV to 200 TeV), a total collection area of $\sim10$ km$^{2}$, and the improved $<0.1^{\circ}$angular resolution, CTA will be about a factor of 10 more sensitive than the current IACTs \cite{2013APh....43..171B} and orders of magnitude better than the current space-based gamma-ray tracking telescopes for short exposures ($<$1 hour) in the overlapping $30-100$ GeV energy range \cite{2013APh....43..348F}.
\\
Given the flaring and transient nature of many VHE sources, the CTA observatory will couple the unprecedented sensitivity to another performance driver: the fast reaction to VHE events, achieved by means of the Real-Time Analysis (RTA) science alert system \cite{2013arXiv1307.6489B}.
The RTA pipeline, one of the products of the CTA On-Site Analysis (OSA, \cite{ABthis}), will automatically detect sub-minute emission and generate science alerts with a latency of 30 seconds with respect to the triggering event collection, allowing both rapid telescope re-pointings and communications (e.g. Virtual Observatory alerts) to external observatories, and making CTA a crucial instrument for the observation of the gamma-ray flaring sky. In addition to a better understanding of the VHE time-variable nature of galactic (e.g. pulsars) and extragalactic (e.g. blazars) sources, the RTA will be a key system for the CTA follow-up study of multi-wavelength and multi-messenger transient sources (e.g. Gamma-Ray Bursts (GRB), Gravitational Waves (GW), VHE neutrinos), and for the discovery of unexpected serendipitous events. For example, it would allow a fast localization of GWs progenitors or the quick re-pointing of additional telescopes on a GRB.

\section{The sensitivity to transient events}

The RTA receives array (or sub-array) events and, after calibration and shower image reconstruction, performs a variability analysis using different integration windows, from seconds to minutes and hours. According to the CTA design requirements, if a transient source is detected in the scanned time-scale, not only the science alert must be sent within 30 seconds from the last acquired event, but the sensitivity (i.e. the minimum detectable flux) to the VHE emission can not be worse than a factor of three the nominal CTA sensitivity.
At the end of the observation, or the morning after, the so-called OSA/Level B analysis pipeline takes place, a more refined analysis system - the sensitivity must be a maximum factor two worse than the nominal one - that produces science monitoring results with a latency of 10 hours.
\\
Each transient target is characterized by a specific spectral behavior, time-variability, intrinsic luminosity, and visibility interval, that define which telescopes are mandatory in the observation run, the required exposure and sensitivity to be detected.
At the same time the CTA sensitivity is defined by (i) the exposure, (ii) the energy-dependent effective area, also directly dependent on the array configuration and exposure time, (iii) the analysis pipeline ability to reject the background, (iv) the angular resolution, (v) the field of view, (vi) the energy threshold and resolution. A detailed trade-off study between the CTA high level science projects and the RTA technological performance is mandatory \cite{AZthis} to define the RTA applicability and feasibility range (i.e. the minimum background rejection efficiency, the minimum exposure).
\\
Once the OSA framework and algorithms will be defined and operative, the RTA sensitivity will be the direct product of the analysis pipeline.
As a first, fundamental step of the trade-off process, the RTA sensitivity is analytically computed here on the basis of the official CTA Monte Carlo (MC) simulations \cite{2013APh....43..171B} for a set of short exposures.

\subsection{Analytical sensitivity evaluation}\label{sec:anal}

\begin{figure}[h!]
     \includegraphics[width=1.\textwidth]{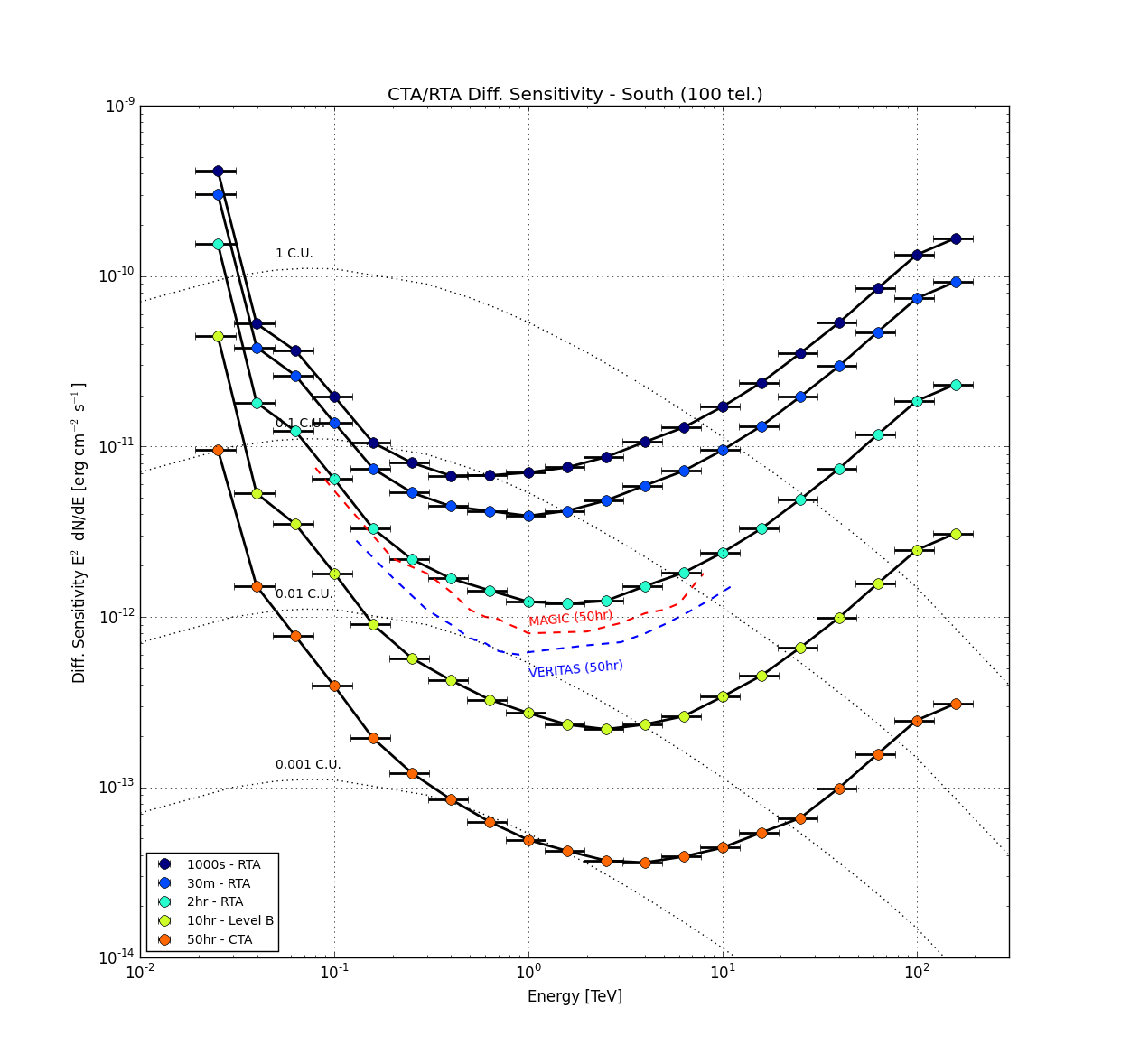}
     \caption{Differential sensitivity, in erg cm$^{-2}$ s$^{-1}$, of the Real-Time and Level B analysis pipelines computed for t$_{\rm exp}$ = 1000s, 30m, and 2hr (RTA) and 10hr (Level B). The expected CTA sensitivity is also shown for t$_{\rm exp}$ = 50hr as reference. 
The dotted lines show the Crab flux in 0.001, 0.01, 0.1, 1, and 10 C.U., while 
the dashed lines show the MAGIC and VERITAS sensitivity for a t$_{\rm exp}$ = 50hr exposure.}
     \label{fig:sens}
\end{figure}

All the results refer to the full CTA Southern array, composed by 4 large-sized telescopes, 24 medium-sized telescopes and 72 small-sized telescopes, for a total covered area of $\sim4$ km$^{2}$.
\\
Since the effective area and the background rate also depend on the simulated exposure, because different algorithms and selection cuts are applied for the background rejection and efficiency computation, dedicated MC campaigns are needed for each exposure under study.
The MC simulation assumes a source with a power law spectrum and a Crab-like index located at the center of the field of view and observed at a zenith angle of 20 degrees. The following full CTA Southern array exposures have been simulated: 1000 seconds, 30 minutes, 2, 10, and 50 hours.
%
%
%
%
%
%
%
%Only three exposures are currently simulated in the official CTA performance page: 50 hours, 5 hours, and 30 minutes. Since the effective area and the background rate also depend on the simulated exposure, because different algorithms and selection cuts are applied for the background rejection and efficiency computation, the very short RTA exposures (from 10 seconds to minutes) will require a dedicated MC simulation campaign. In the present work, we assume a constant effective area and background rate from 1000 seconds to 1 hour, and from 3 to 10 hours, using the 30 minutes and 5 hours integrated MC simulation simulations respectively.
\\
Generally, the minimum detectable flux in the gamma-ray energy range is given by the required confidence level for detection and the statistics of the detected photons. Each energy decade is divided in 5 energy bins, with equal logarithmic bin width. Following the rules of the CTA performance evaluation \cite{2013APh....43..171B}, the differential sensitivity is computed by requiring, for each energy bin, a statistical significance of 5 standard deviations ($\sigma$) of the gamma-ray excess above the background. The significance is obtained using the equation (17) of \cite{1983ApJ...272..317L}, assuming a ratio of the on-source to the off-source exposure time $\alpha=0.2$ (i.e. the background emission is taken from a 5 times larger region than the gamma-ray event region). A minimum number of 10 source photons must be collected to gain a detection, with an excess above the background level greater than 5\%.
\\
The time latency constraints in the RTA science alert production and the Level B general science monitoring could impose the use of, e.g., unrefined selection cuts and conservative significance requirements. For this reason, and following the CTA design, a sensitivity decreasing factor of three and two is applied. However we remind the reader that fast multivariate analysis (MVA) methods, as those currently applied in the latest CTA MC results, could speed-up the RTA pipeline to achieve the full CTA sensitivity within the real-time latency constraints.
\\
The RTA minimum energy flux, in E$^{2}$ dN/dE, is computed for the following t$_{\rm exp}$ intervals: 1000 seconds, 30 minutes, 2 hours. The Level B analysis sensitivity is computed for a 10 hours long observational night, while the CTA full sensitivity is presented for the standard t$_{\rm exp} = 50$ hours. The results are shown in Fig. \ref{fig:sens}, with the exposure time increasing from top to bottom. 
The fractional Crab flux, in Crab units (C.U.), is superimposed using the model of  \cite{2010A&A...523A...2M}. 
The differential sensitivity of the current MAGIC and VERITAS telescopes (\cite{CTA1} and references therein), for an integration window of 50 hours are also shown. 

\section{Results and discussions}\label{sec:res}

The main result from this preliminary RTA performance evaluation is that in few hours of observation the southern full CTA array will collect the same amount of photons a 50 hours of integration of the current MAGIC and VERITAS telescopes while being able to detect and send science alerts within 30 seconds, as shown in Fig. \ref{fig:sens}. In addition, with the same few hours of observation CTA will be capable of observing Crab-like sources in the unexplored $>10$ TeV energies.
\\
Considering that the duration of the VHE extended emission of short GRBs associated to GW progenitors is expected to reach 1000 seconds \cite{2014MNRAS.443..738B}, we select this exposure as the value of reference for the present analysis. For the three selected energies {0.05, 0.1, 1} TeV, and an exposure t$_{\rm exp} = 1000$ seconds, we obtain - in the worst case - an RTA differential sensitivity of {$6.2\times10^{-11}$, $1.9\times10^{-11}$, $5.2\times10^{-12}$} erg cm$^{-2}$ s$^{-1}$, respectively, or a fluence, in 1000 seconds, of {$62\times10^{-9}$, $19\times10^{-9}$, $5.2\times10^{-9}$} erg cm$^{-2}$.  Assuming a short GRB pointed by CTA after 30 seconds from the emission start (E$_{\rm kin}=10^{51}$ erg, D$_{\rm L}$=300 Mpc), according to \cite{2014MNRAS.443..738B} the following fluence, for t$_{\rm exp} = 1000$s, is expected: $360\times10^{-9}$, $310\times10^{-9}$, $170\times10^{-9}$ erg cm$^{-2}$, for {0.05, 0.1, 1} TeV. Even adopting a worsening factor of three in sensitivity, the CTA/RTA pipeline will be capable to trigger this class of GRBs. 
At 1 TeV, a short GRB captured after 1000 seconds from the start of the emission could also be observed.
\\
The CTA energy coverage can be divided in three main energy ranges (0.01 - 0.1 TeV, 0.1-10 TeV, and $>10$ TeV) where background fluctuations, background rate and effective area play a dominant role. At low energies, the night sky and especially the cosmic-ray induced  background is extremely high, inducing systematic errors in the background subtraction \cite{2007A&A...466.1219B}. In particular, the sensitivity is limited by cosmic-ray electrons between 100 GeV and 800 GeV, for long exposures. Around 1 TeV, the sensitivity follows the background statistics induced by cosmic-ray protons that convert most of their energy into electromagnetic cascades in the first interactions, causing "gamma-like" showers that are difficult to cut \cite{1999APh....11..247V}. At higher energies the sensitivity is signal limited, because the occurrence of VHE photons naturally decreases and the detection is directly dependent on the total effective area.
\\
The use of shorter exposures has less impact on the sensitivity below < 100 GeV, with a sensitivity loss about 2 times lower than what occurs at 1 TeV if we compare the 3 hr to the 1000 s exposures. At the same time, for shorter exposures the $>1$ TeV energy regime is less influenced by the background rate.
%higher is the energy and lower is the influence of the background rate on the sensitivity
%the signal-limited energy range shows a decrease in the lower energy threshold: as the exposure decreases, the background has a less impact on the VHE event detection. 
%For this reason, the use of the maximum number of small-size telescopes, i.e. the maximum collection area, is the key to a better reaction from flaring events if short exposures are mandatory. 
However, because of the steep power-law shape of many VHE sources and the absorption by the Extragalactic Background Light for high z sources, follow-up observations of many VHE transient sources, including GRBs \cite{2013APh....43..252I}, require the use of the large-sized, low energy, telescopes, where the sensitivity is less dependent on the exposure time and sub-arrays could be used.
\\
%The main result from this preliminary RTA performance evaluation is that the CTA observatory will be able, for the first time in the TeV energy range, to map the fast variability of many transient phenomena with an unprecedented sensitivity.: in few hours of observation CTA not only will collect the same amount of photons of 50 hours of integration of the current MAGIC and VERITAS telescopes, it will also be capable to send science alerts within 30 seconds. Below 1 TeV, it will observe in less than 30 minutes the same source of a 5 years HAWC integration. 
%The CTA official science requirements define a maximum differential sensitivity of $8\times10^{-12}$, $2\times10^{-13}$, and $3\times10^{-13}$ erg cm$^{-2}$ s$^{-1}$ at 50 GeV, 1 TeV, 50 TeV (see Fig. \ref{fig:sens}). The CTA nominal differential sensitivity obtained from the last MC simulation campaign is well beyond the requirement, which can be even satisfied in 10 hours of Level B analysis, especially below 10 TeV.
Although the differential sensitivity allows the characterization of the spectral shape, the integral sensitivity, i.e. the minimum detectable flux above a certain energy threshold, defines the CTA ability to detect new VHE source. The computation of the RTA integral sensitivity for a set of science target spectra is planned in the near future.
%Thanks to the wide energy coverage, the integral sensitivity could even be better. 
\\
%\subsection{Impact on the technological requirements}
The analytical sensitivity evaluation presented here uses the background rate resulting from the standard off-line CTA analysis. However, different algorithms and techniques could be used by the OSA pipeline. At the same time, we assume that the full Southern array is deployed in the observation, but sub-arrays covering different energy ranges will be used in many cases. 
\\
Waiting for dedicated MC simulations of the CTA sub-arrays at short exposures, and as a preliminary trade-off study of the RTA performance definition, we could analytically explore the feasibility range of two important requirements: the background rate and the effective area. 
While above 1 TeV, for short exposures, the ability to detect VHE sources is mainly defined by the number of small-sized telescopes pointing the sources, at lower energies both the effective area and the background rejection efficiency contribute to the CTA sensitivity. 
\\
Keeping as primary goal the maximization of the science return by an efficient observational scheduling of the CTA array, given the science target and its typical flux and spectral shape, the impact of different array configuration and background rejection techniques on the final sensitivity will be studied in the near future.% In particular, for a selected sensitivity value, the dependency of the background rate on the effective area variations will be explored in detail.
\\
%Although Fig. \ref{fig:trade} reports the background rate, in Hz, to be compared to the CTA official performance, the background rate should be divided by the input effective area to obtain an absolute background flux.
%Table \ref{tab:trade} lists the background rate required to achieve the same sensitivity 
%to divide in half the 50 and 100 GeV effective area by, e.g., reducing the number of large-sized telescopes or changing the telescope distance. At 1 TeV, 2/3 of the effective area are used. Given the general assumptions and the analytical evaluation, we can affirm that the use of half effective area is possible if the selection cuts are able to reduce of about 70\% the nominal background rate. 
\\
%Although very preliminary, the present study shows that the connection between different parameters can be analyzed to define the best scenarios for the RTA use cases and the maximization of the science return. 

\section{Summary and conclusions}
The CTA Consortium scheduling plan is approaching the construction phase \cite{CTA1}, and scientific data taking will start as soon as the first telescopes will be in place. For this reason, the evaluation of the CTA On-Site Analysis performance as a function of the array configuration (e.g. number or type of telescopes) and observational requirements (e.g.  time resolution and duration, minimum flux, zenith angle) of the flaring VHE sources is mandatory for the preparation of the scientific use cases.
\textbf{Assuming a factor three of sensitivity decrease respect with the nominal CTA performance, the RTA pipeline, with the southern full array, will achieve in few hours of observations and within 30 seconds of reconstruction and analysis processing the same sensitivity of 50 hours of integration of the current Cherenkov telescopes.} 
However, the future improvement in the analysis algorithms and processing hardware could greatly increase the OSA pipeline performance, and the real-time minimum detectable flux could approach the nominal CTA sensitivity.
\\
These results establish the role of CTA as the observatory of reference for the TeV real-time study of the flaring sky, not only for the discovery of serendipitous sources in the field of view but also for follow-up studies from external triggers. The definition of the CTA observational strategy for transient sources will depend on each specific target and the array configuration. 
%A preliminary and analytical trade-off study of the background rate and effective area impact on the final sensitivity shows that at low energies the use of sub-arrays guarantees the scientific requirements, if the background rejection efficiency is increased of more than 70\%. 
\\
In the near future, the use of sub-array MC simulations and actual reconstruction algorithms, coupled with the evaluation of a set of reference spectra and light curves of transient science targets, will allow the detailed definition and optimization of the RTA science cases.

\acknowledgments
VF would like to thank Gernot Maier for the fruitful discussion. We gratefully acknowledge support from the agencies and organizations 
listed under Funding Agencies at this website: http://www.cta-observatory.org/.

\end{document}